\documentclass[nofootinbib,prd,twocolumn,showpacs,showkeys,preprintnumbers]{revtex4-1}
\usepackage{hyperref,amssymb,amsmath,mathrsfs,bm,graphicx}
\usepackage{multirow}
\usepackage{array}
\usepackage{dblfloatfix}
\usepackage{dblfloatfix}
\usepackage{placeins}
\usepackage{float}
\usepackage[dvipsnames]{xcolor}
\begin{document}

\title {Construction of a traversable wormhole from a suitable  embedding function.}

\author{R. \'Avalos}
\affiliation{Departamento de F\'isica, Colegio de Ciencias e Ingenier\'ia, Universidad San Francisco de Quito,  Ecuador.\\}
\author{E. Fuenmayor }
\affiliation{Centro de Física Teórica y Computacional, Escuela de Física, Facultad de Ciencias, Universidad Central de Venezuela, Caracas 1050, Venezuela.\\}

\author{E. Contreras}
\email{econtreras@usfq.edu.ec}
\affiliation{Departamento de F\'isica, Colegio de Ciencias e Ingenier\'ia, Universidad San Francisco de Quito,  Quito 170901, Ecuador.\\}

\begin{abstract}
In this work we construct traversable wormholes geometries in the framework of the complexity factor. We provide the redshift function of a Casimir traversable wormhole which, in combination with a non--vanishing complexity factor, leads to a traversable wormhole with a minimum amount of exotic matter. The shape function and the embedding diagram are shown and discussed. The tidal accelerations and the time required to get through the wormholes are estimated. 
\end{abstract}
\maketitle
\section{Introduction}
The simple thought of rapid interstellar travels is a 
intriguing idea and wormholes seems to be an alternative to achieve such a goal (at least theoretically). The idea of wormholes can be traced back to the work of Flamm  \cite{flamm} who recognized that the Schwarzschild black hole could provide a way for interstellar travels. However, the huge tidal forces and the dynamical nature of the throat forbids its use as a humanly traversable wormhole.\\

The basis of traversable wormholes were stated in the seminal work of Morris and Thorne \cite{thorne}. In this work the authors construct the desired geometry which could provide a bridge between asymptotically flat regions of the same universe (or connect two different universes) and then ask for the matter content which is required to maintain such a tunnel open with the aim to be used for interstellar travels. The result is that the required matter content necessarily violates the null energy condition (NEC) so that construction of traversable wormholes depends on the existence of such an ``exotic matter'' in the universe. Although there is not evidence for the existence of exotic matter, it is expected that some exotic field or quantum states of known fields, violating NEC on macroscopic length scales, allow the construction of traversable wormholes. For example, in \cite{lobo} it has been suggested the existence of wormholes supported by phantom energy and beside, in Refs. \cite{thorne2,visserCas,lobo2,remo} it has been proposed that the Casimir effect could provide such an artificial source of exotic matter realizable in a laboratory. In this regard, given the (theoretical) feasibility of wormholes, the seeking for such geometries is still an active research line 
(see \cite{visserCas,lobo} for a self--contained discussion on wormhole. Also see \cite{Chojnacki:2021xtr,Baruah:2021fge, Kundu:2021nwp,Visser:2021vkt,Stuchlik:2021guq,Bronnikov:2021ods,Blazquez-Salcedo:2021udn,Bronnikov:2021piw,Churilova:2021tgn,Konoplya:2021hsm,Tello-Ortiz:2021kxg,Bambi:2021qfo,Sarkar:2021uob, Capozziello:2020zbx,Blazquez-Salcedo:2020czn, Berry:2020tky,Maldacena:2020sxe, Garattini:2020kqb}, for recent developments). 

From a more technical point of view, the 
construction of traversable wormholes requires to find solutions of the Einstein field equations with a very specific set of boundary conditions and some extra information to close the system of differential equations. For example, in static and spherically symmetric spacetimes this corresponds to solve a set of three coupled differential equations with five unknowns (two metric functions, the energy density and the radial and tangential pressures) so that, besides the boundary conditions which ensure the traversable wormhole geometry, two extra constraints must be supplied. A possible strategy to close the system is providing two metric functions that fulfill the required boundary conditions in order to obtain the matter content (indeed, this was the strategy followed in \cite{thorne}). Other possibilities could be either to supply an equation of state and one of the metric functions or to provide one of the metric functions and a metric constraint. In this work we propose an alternative route consisting in setting of a suitable metric and a specific complexity factor introduced in \cite{complex1} which has been broadly used as a complementary condition to solve the Einstein equations in different contexts \cite{Yousaf:2020dci,Abbas:2018idr,Abbas:2020qzz,Herrera:2019cbx,Sharif:2021gsl,Sharif:2021fdj, Annals}. The reason of introducing the complexity factor is twofold. First, it provides a natural measurement of complexity at macroscopic scales \cite{complex1} which entails a non-local equation of state relating both the density contrast and the local anisotropy of pressure through a simple scalar arising from the orthogonal decomposition of the Riemann tensor \cite{Bel,Garcia,EF}. Second, each value of the so called complexity factor allow to define a kind of equivalence class of solutions of Einstein field equations: two solutions with the same complexity factor belong to the same class.\\

This work is organized as follows. In the next section we introduce the basic aspects of traversable wormholes. In section \ref{complex} we review the main aspects of the complexity factor. Then, in section \ref{TWC} we explore how to construct wormhole geometries imposing certain non--vanishing complexity. The last section is dedicated to the conclusions and final remarks of the work.

\section{Traversable wormholes}

In this section we summarize the main properties that a solution of Einstein field equations must satisfy in order to describe a traversable wormhole. The first property is that the geometry must provide a throat connecting two asymptotically flat regions. Besides, any horizon should be absent if the two--way travel is desired.  Another property is that, in order to be humanly traversable, the tidal forces experienced by a traveler must be ``small'' (bearable by a human being). Additionally, the proper time to traverse the wormhole should be finite. However, all of these geometric properties listed above constraint the matter content in such a manner that the energy conditions are violated as we demonstrate in what follows.

Let us consider a line element parametrized as
\begin{eqnarray}\label{metric}
ds^{2}=-e^{2\phi} dt^2 +d r^{2}/(1-b/r)+r^{2}(d\theta^{2}+\sin^{2}\theta d\phi^{2}),\nonumber\\
\end{eqnarray}
where $\phi$ and $b$ are the so called redshift and shape functions respectively and are functions of the radial coordinate only. We 
assume that (\ref{metric}) is a solution of Einstein's equations
\begin{eqnarray}\label{EFE}
R_{\mu\nu}-\frac{1}{2}g_{\mu\nu}R=\kappa T_{\mu\nu},
\end{eqnarray}
with $\kappa=8\pi G/c^{2}$ \footnote{In this work we shall assume $c=G=1$.}, sourced by 
$T^{\mu}_{\nu}=diag(-\rho,p_{r},p_{t},p_{t})$. Using (\ref{metric})
the Einstein's field equations (\ref{EFE}) read
\begin{eqnarray}
\rho&=&\frac{1}{8\pi}\frac{b'}{r^{2}}\label{rho}\\
p_{r}&=&-\frac{1}{8\pi}\left[\frac{b}{r^{3}}-2\left(1-\frac{b}{r}\right)\frac{\phi'}{r}\right]\label{pr}\\
p_{t}&=&\frac{1}{8\pi}\left(1-\frac{b}{r}\right)
\bigg[\phi''+(\phi')^{2}-\frac{b'r-b}{2r^{2}(1-b/r)}\phi'\nonumber\label{pt}\\
&&-\frac{b' r-b}{2r^{3}(1-b/r)}+\frac{\phi'}{r}\bigg].
\end{eqnarray}

As there is not horizon, $g_{tt}$ must be a non vanishing function to avoid the existence of a infinite redshift surface, then 
$\phi$ must be finite everywhere.\\

The information about the throat of the wormhole is encoded in its
shape. First, given that our solution is spherically symmetric, let us consider $\theta=\pi/2$ without loss of generality. Next, considering a fixed time, $t=constant$, the line element reads
\begin{eqnarray}\label{emb1}
ds^{2}=\frac{dr^{2}}{1-b/r}+r^{2}d\phi^{2}.
\end{eqnarray}
Note that this surface can be embedded in a three dimensional space in which the metric can be written in cylindrical coordinates $(r,\phi,z)$ as
\begin{eqnarray}
ds^{2}=dz^{2}+dr^{2}+r^{2}d\phi^{2}.
\end{eqnarray}
Now, as $z$ is a function of the radial coordinate we have
\begin{eqnarray}
dz=\frac{dz}{dr}dr,
\end{eqnarray}
from where
\begin{eqnarray}\label{emb2}
ds^{2}=\left[1+\left(\frac{dz}{dr}\right)^{2}\right]dr^{2}
+r^{2}d\phi^{2}.
\end{eqnarray}
After combining (\ref{emb1}) and (\ref{emb2}) we obtain
\begin{eqnarray}\label{emb3}
\frac{dz}{dr}=\pm 
\left(\frac{r}{b}-1\right)^{-1/2},
\end{eqnarray}
from where is clear that $b>0$ for $r\in[r_{0},\infty)$.
The requirement of a throat entails that the wormhole geometry must be endowed with a radius $r=b_{0}$ where the $r(z)$  is a minimum, namely $dz/dr\to\infty$ as $r\to b_{0}$ and from (\ref{emb3}) it occurs when $b=r$. In this regard, the
existence of a minimum
radius requires that at $r=b_{0}$ the shape function must be $b=b_{0}$. Additionally we demand that the solution is asymptotically flat which implies both, $b/r\to0$ (from where $dz/dr\to0$) and $\phi\to0$ as $r\to\infty$. 

It is worth noticing that as the conditions 
\begin{eqnarray}
&&\lim\limits_{r\to b_{0}}\frac{dz}{dr}\to\infty\\
&&\lim\limits_{r\to\infty}\frac{dz}{dr}=0,
\end{eqnarray} 
must be satisfied, the smoothness of the geometry is ensured whenever the embedding surface flares out at or near the throat. To be more precise, as $dr/dz=0$ (a minimum) at the throat, we impose 
\begin{eqnarray}
\frac{d^{2}r}{dz^{2}}>0,
\end{eqnarray}
from where
\begin{eqnarray}
\frac{b-b'r}{2b^{2}}>0,
\end{eqnarray}
which corresponds to the 
flaring--out condition. Besides ensuring
the expected behaviour of a traversable wormhole, the flaring--out condition leads to a constraint as we shall see in what follows. Let us define the quantity
\begin{eqnarray}
\xi=-\frac{p_{r}+\rho}{|\rho|}=\frac{b/r-b'-2(r-b)\phi'}{|b'|},
\end{eqnarray}
which can be written as
\begin{eqnarray}
\xi=\frac{2b^{2}}{r|b'|}\frac{d^{2}r}{dz^{2}}
-2(r-b)\frac{\phi'}{|b'|}
\end{eqnarray}
Now, as $(r-b)\to0$ at the throat, we have 
\begin{eqnarray}
\xi=\frac{2b^{2}}{r|b'|}\frac{d^{2}r}{dz^{2}}>0
\end{eqnarray}
so that
\begin{eqnarray}
\xi=-\frac{p_{r}+\rho}{|\rho|}>0.
\end{eqnarray}
Note that if $\rho>0$ the above condition implies $p_{r}<0$ which entails that $T^{1}_{1}$ should be interpreted as a tension. Furthermore, if we define $\tau=-p_{r}$ the flaring out condition leads to
\begin{eqnarray}
\tau-\rho>0,
\end{eqnarray}
which implies that the throat tension must be greater than the total energy density which violates the null energy condition (NEC).
The kind of matter satisfying such a condition is called exotic matter. However, although there is not evidence of exotic matter in the universe, we can 
minimize the amount required to construct a traversable wormhole by demanding that the integral (known as quantifier) \cite{visserQ}
\begin{eqnarray}\label{visserQ}
I=\int dV (\rho+p_{r})=-\int\limits_{r_{0}}^{\infty}
(1-b')\left[\ln \left(\frac{e^{2\phi}}{1-b/r}\right)\right] dr \nonumber\\
\end{eqnarray}
is finite.\\

In order to obtain a wormhole that can be humanly traversable we must ensure that, in the acceleration experience by the traveler,
the radial and lateral tidal constraints must be bounded as
\begin{eqnarray}
&&\bigg| \bigg( 1-\frac{b}{r} \bigg) \bigg[  \phi'' + (\phi ')^2 -\frac{r b' -b}{2r(r-b)} \phi '\bigg]  \bigg|| \eta^1|\le\frac{g_{\oplus}}{c^{2}}\label{tidalr}\\
&&\bigg| \frac{\gamma^2}{2r^{2}}\left[\beta^{2}(b'-\frac{b}{r})+2r(r-b)\phi'\right]\bigg||\eta^{2}|\le\frac{g_{\oplus}}{c^{2}}\label{tidalt},
\end{eqnarray}
with $g_{\oplus}$ the Earth's gravitational acceleration, $\gamma=1/(1-\beta^{2})^{1/2}$, $\beta=v(r)/c$, $c$ the speed of light and
$\eta^{1}$ and $\eta^{2}$ the radial and lateral size of the traveler respectively. 

Finally, to ensure that the trip is completed in a reasonable time, we demand that the coordinate and proper time are bounded as
\begin{eqnarray}
\Delta t&=&\int\limits_{r_{0}}^{r_{st}}\frac{e^{-\phi}dr}{v\sqrt{1-b/r}}<1\  \textnormal{year},\label{coort}\\
\Delta\tau&=&\int\limits_{r_{0}}^{r_{st}}\frac{dr}{v\gamma\sqrt{1-b/r}}<
1\ \textnormal{year}\label{propt},
\end{eqnarray}
where $r_{st}$ is the radial coordinate that represents the station location.

From a technical point of view, the construction of traversable wormholes require solving the system (\ref{rho})-(\ref{pt}), namely three equations with 
five unknowns.  The strategy should be either supplying the metric functions that satisfy the geometric constraints listed above or giving one of the metrics and an auxiliary condition, namely an equation of state or a metric constraint like the class I (Karmarkar) \cite{karmarkar} or the vanishing Weyl's tensor (conformally flat) condition \cite{weyl}, for example. In this work we follow an alternative route which consists of proposing a suitable metric function and
the complexity factor ($Y_{TF}$ given in \cite{complex1} below) as an auxiliary condition which shall be introduced in the next section.

\section{Complexity Factor}\label{complex}

Recently, a new definition for complexity for self--gravitating
fluid distributions has been introduced in
Ref. \cite{complex1}. This definition is based on the intuitive idea that the least complex gravitational system should be characterized by a homogeneous energy density distribution with isotropic pressure. Now, as demonstrated in \cite{complex1}, there is a scalar associated to the orthogonal splitting of the Riemann tensor \cite{Garcia, EF} in static spherically symmetric space--times that captures the essence of what we mean by complexity, namely
\begin{eqnarray} \label{YTF2}
Y_{TF} = 8\pi \Pi - \frac{4\pi}{r^3}\int\limits_{0}^{r} \tilde{r}^3 \rho' d\tilde{r},
\end{eqnarray}
with $\Pi\equiv p_{r}-p_{\perp}$. Also, it can be shown that (\ref{YTF2}) allows to write the Tolman mass as,
\begin{eqnarray} \label{m_T}
m_{T} = (m_{T})_{\Sigma}\left(\frac{r}{r_{\Sigma}}\right)^3 + r^3\int^{r_{\Sigma}}_{0} \frac{e^{( \nu + \lambda )/2}}{{\tilde{r}}} Y_{TF} d\tilde{r},
\end{eqnarray}
which can be considered as a
solid argument to define the complexity factor by means of this scalar given that this function, encompasses all the modifications produced by the energy density inhomogeneity and the anisotropy of the pressure on the active gravitational mass. \\

Note that the vanishing complexity condition ($Y_{TF}=0$) can be satisfied
not only in the simplest case of isotropic and homogeneous system but in all the cases where
\begin{eqnarray}
\Pi=\frac{1}{2r^{3}}\int\limits\tilde{r}^{3}\rho' d\tilde{r}.
\end{eqnarray}
In this respect, the vanishing complexity condition leads to a non--local equation of state that can be used as a complementary condition to close the system of EFE (for a recent implementation, see \cite{Annals, casadioyo, EC-EF}, for example). 

In this work, we are interested in to construct traversable wormholes by using the definition of complexity factor as a non--linear equation of state. However, as traversable wormholes are defined for $r\in[r_{0},\infty)$, the $Y_{TF}$ given by Eq. (\ref{YTF2}) should be modified properly. To be more precise, the complexity factor for traversable wormholes must be defined as
\begin{eqnarray} \label{YTF22}
Y_{TF} = 8\pi \Pi - \frac{4\pi}{r^3}\int\limits_{r_{0}}^{r} \tilde{r}^3 \rho' d\tilde{r},
\end{eqnarray}
which leads to
\begin{eqnarray} \label{YTF23}
&&Y_{TF}=\bigg(1-\frac{b}{r}\bigg)\bigg(\frac{\phi'}{r}- \phi'^2  - \phi ''\nonumber\\
&&+\phi' \bigg(1-\frac{b}{r}\bigg)^{-1} \bigg( \frac{r b'-b}{2r^2} \bigg)
\bigg)+\frac{r_{0}b'(r_{0})-3r_{0}}{2r^{3}}.
\end{eqnarray}
It is worth noticing that, the standard definition demands $r_{0}=0$ but we must discard this case to ensure a finite size of the wormhole throat.\\

From (\ref{YTF23}) it is straightforward to observe that a traversable wormhole, with a constant redshift function, fulfills the vanishing complexity condition in a trivial way whenever $b'(r_{0})=3$. More precisely, Eq. (\ref{YTF23}) cannot be used as an equation of state to obtain the shape function, and in consequence $b$ is any arbitrary suitable shape function with $b'(r_{0})=3$. However, it is possible to construct a traversable wormhole with vanishing complexity by supplying either a non--constant redshift function or
a suitable shape function. Similarly, we may provide a particular value of $Y_{TF}$ and then use this information to find a family of solutions with the same complexity factor \cite{casadioyo}. In this case, particular values of $Y_{TF}$ allow to define a kind of equivalence class of solutions; namely, two solutions with the same complexity factor are equivalent. In this work, we shall implement both approaches that we have just mentioned in order to construct traversable wormholes based on the metrics of a Casimir wormhole \cite{remo}, as we shall explain in the next section.

\section{Traversable Casimir wormhole}\label{TWC}

In Ref. \cite{remo} it has been constructed a traversable wormhole supported by a matter sector satisfying
\begin{eqnarray}
p_{r}&=&\omega\rho ,\\
p_{t}&=&\omega_{t}(r)\rho,
\end{eqnarray}
with $\omega$ a constant and
\begin{eqnarray}
\rho&=&-\frac{r_{0}^{2}}{8\pi\omega r^{4}},\\
\omega_{t}(r)&=&-\frac{\omega^{2}(4r-r_{0})+r_{0}(4\omega+1)}{4(\omega r+1)},
\end{eqnarray}
where $r_{0}\approx 1.016l_{P}$ the size of the throat.\footnote{$l_{P}$ is the Planck size so the solution corresponds to a traversable wormhole of Planck size.} The metric functions for this solution reads
\begin{eqnarray}
\phi&=&\frac{1}{2}(\omega-1)\ln\left(\frac{r\omega}{\omega r+r_{0}}\right),\label{rs1}\\
b&=&\left(1-\frac{1}{\omega}\right)r_{0}+\frac{r_{0}^{2}}{\omega r}\label{sh1}.
\end{eqnarray}
It is noticeable that if $\omega=3$ the equation of state for the radial pressure has the Casimir form ($p_{r}=3\rho$) but the entire matter sector has an extra contribution due to tangential pressure.

For $\omega=3$ some interesting features should be highlighted. First, assuming a constant speed for the traveler and $\gamma\approx1$, the acceleration leads to
\begin{eqnarray}
|a|=|\sqrt{1-\frac{2r_{0}}{3r}-\frac{r_{0}^{2}}{3r}}\frac{r_{0}}{r(3r+r_{0})}|\le\frac{g_{\oplus}}{c^{2}},
\end{eqnarray}
from where it can be seen that at the throat the traveller has a vanishing acceleration. Then, the radial and lateral tidal constraints near to the throat are given by
\begin{eqnarray}
r_{0}&\gtrsim&10^{8}m ,\\
v&\lesssim& 2.7 r_{0}s^{-1}.
\end{eqnarray}
Next, if the traveler journeys with constant speed $v$ both the coordinate and the proper time are of the order of $\sim 5\times 10^{3}s$, assuming that the station is located at $r_{st}=10^{4}r_{0}$. Finally, the quantifier for this solution is
\begin{eqnarray}
I=-\frac{4r_{0}}{\kappa}
\end{eqnarray}
which could be arbitrarily small depending on the values of $r_{0}$.\\

In the next section we construct a wormhole geometry by assuming a generalization of both the redshift function (\ref{rs1}) and the complexity factor of the solution with $\omega=3$. As we shall see, the resulting shape function will have a set of free parameters which we fix by demanding the conditions for a humanly traversable wormhole.

\section{Traversable wormholes and complexity}\label{TWC}
Setting $\omega=3$ (Casimir wormhole) in
(\ref{rs1}) and (\ref{sh1}) we obtain
\begin{eqnarray}
\phi&=&\ln\left(\frac{3r}{3r+r_{0}}\right)\label{remo1}\\
b&=&\frac{2r_{0}}{3}+\frac{r_{0}^{2}}{3r}\label{remo2},
\end{eqnarray}
from where the complexity factor reads
\begin{eqnarray}
Y_{TF}&=&-\frac{r_0 (36 r^4+96r^3 r_0 + 16r^2 r^2_0+27 r r^3_0+5 r^4_0)}{6 r^5 (3r+r_0)^2}.\nonumber\\
\label{ytf}
\end{eqnarray}

Then, as a starting point we propose a generalization of both Eqs. (\ref{remo1}) and (\ref{ytf}), namely
\begin{eqnarray}
\phi&=&\ln{\bigg( \frac{c_{0} r}{c_{0} r +r_0} \bigg)}\label{e3}\\
Y_{TF}&=&-\frac{r_0 (a_4 r^4+ a_3 r^3 r_0 + a_2 r^2 r^2_0+ a_1 r r^3_0+ a_0 r^4_0)}{ r^5 (c_{0}r+r_0)^2}\nonumber\\ 
&&+\frac{r_{0}b'(r_{0})-3r_{0}}{2r^{3}}
\label{e4}
\end{eqnarray}
with $c_{0}$, $a_{0}$, $a_{1}$, $a_{2}$, $a_{3}$ and $a_{4}$ constants. Note that as (\ref{e4}) coincides formally with (\ref{ytf}), we shall name it as a like--Casimir complexity factor. Furthermore, every solution for different values of the parameters involved should be considered as ``equivalents'' in the sense that they have the same complexity.

Now, after replacing Eqs. \eqref{e3} and \eqref{e4} in \eqref{YTF23} the term $(r_{0}b'(r_{0})-3r_{0})/2r^{3}$ cancels out and we end with a differential equation for $b$ which solution reads\footnote{The vanishing complexity condition leads to a $b$ which violates all the requirements of a traversable wormhole}
\begin{eqnarray}
b&=&\frac{1}{30r^2 r^5_0} \bigg(12 a_0 r^8_0 +(15 a_1 -27 a_0 c_{0})r^7_0 r \nonumber\\ 
&&+(20a_2 -40 a_1 c_{0} +72 a_0 c_{0}^{2})r^6_0 r^2 \nonumber\\ 
&&+(15+15a_3-35a_2c_{0}+70 a_1 c_{0}^{2}-126a_0 c_{0}^{3})r^5_0 r^3 \nonumber\\
&&+(5 a_4- 15a_3 c_{0} + 35 a_2 c_{0}^{2}-70 a_1 c_{0}^{3} + 126a_0 c_{0}^{4})\nonumber\\
&&\times\big[ 12r^4 r^4_0+125 r^5 r^3_0 +260r^6 r^2_0 +210 r^7 r_0\nonumber\\   
&&+60 r^8 +r^5 (c_{0} r +r_0)^4    \frac{60c_{0}}{r_0} \ln\frac{r}{c_{0} r +r_0} \big] \bigg).
\end{eqnarray}
At this point it is clear that the solution depends on six free parameters that we shall restrict by imposing the conditions that a suitable traversable wormhole has to satisfy. \\

In order to ensure that the solution is asymptotically flat we demand
\begin{eqnarray}
&&15+15a_3-35a_2c_{0}+70 a_1 c_{0}^{2}-126a_0 c_{0}^{3}=0,\\
&&5 a_4- 15a_3 c_{0} + 35 a_2 c_{0}^{2}-70 a_1 c_{0}^{3} + 126a_0 c_{0}^{4}=0,
\end{eqnarray}
from where
\begin{eqnarray} 
    a_4&=&-3 c_{0},\\
    a_3&=&-1+\frac{7}{3} a_2 c_{0} - \frac{14}{3} a_1 c_{0}^{2} +\frac{42}{5} a_0 c_{0}^{3},
\end{eqnarray}
and as a consequence,
\begin{eqnarray} \label{e5}
b&=&\frac{1}{30r^2} \bigg( 12 a_0 r^3_0 +(15 a_1 -9 a_0 c_{0})r r^2_0\nonumber\\
&&+ (20a_2 -40 a_1 c_{0} +72 a_0 c_{0}^{2}) r^2 r_0 \bigg).
\end{eqnarray}
A further restriction on the parameters is obtained by imposing $b(r_0)=r_0$ which leads to
\begin{equation} \label{e6}
    a_2=\frac{3}{2}-\frac{3 a_0}{5}\bigg(1-\frac{9c_0}{4}+6c_{0}^{2}\bigg)-a_1\bigg(-\frac{3}{4}-2c_{0}\bigg),
\end{equation}
from where
\begin{eqnarray} \label{e11}
    b&=&(1-\frac{1}{2}a_1 -\frac{2}{5}a_0 +\frac{9}{10} a_0 c_{0}) r_{0}\nonumber\\
    &&\hspace{30pt}+ (\frac{1}{2}a_1 - \frac{9}{10} a_0 c_{0})\frac{r_{0}^2}{r}  + \frac{2}{5} a_0\frac{r_{0}^3}{r^2} .
\end{eqnarray}

Now, the flaring--out condition at the throat 
demands $1>b'(r_{0})$ from where
\begin{equation} \label{e9}
    a_1>-\frac{8}{5}a_0 +\frac{9}{5}a_0 c_{0} -2.
\end{equation}

Although the above conditions are the basic requirements for a traversable wormhole we can restrict the parameter space further if we demand that the geometry is sustained by a finite quantity of exotic matter. In this regard, we have to compute the quantifier given by Eq. (\ref{visserQ}). Replacing (\ref{rho}) and (\ref{pr}) in (\ref{visserQ}) and using (\ref{e3}) and (\ref{e11}) we obtain
\begin{eqnarray} \label{q3}
I&=&\int^{\infty}_{r_0} \Bigg[\frac{r_0}{(c_{0} r+r_0)}  \bigg( 1-\frac{c_{0}}{2}+\frac{a_0 c_{0}}{5} +\frac{a_1 c_{0}}{4} -\frac{9 a_0 c_{0}^2}{20} \bigg)\nonumber\\
&&-\frac{a_0 r_0^4}{r^3 (c_{0} r +r_0)} + \frac{r_0^3}{r^2 (c_{0}r+r_0)} \bigg(-a_1 +\frac{6 a_0c_{0}}{5} \bigg)
 \nonumber\\
&&
+\frac{r_0^2}{r(c_{0} r +r_0)} \bigg(-\frac{3}{2}+\frac{3a_0}{5} +\frac{3a_1}{4}\nonumber\\ &&-\frac{27a_0c_{0}}{20} -\frac{a_1c_{0}}{2}+\frac{9a_0c_{0}^{2}}{10} \bigg)
\Bigg] dr. \label{I}
\end{eqnarray}
Note that $I$ converge whenever
the first term in (\ref{q3}) vanish, namely
\begin{eqnarray}
 1-\frac{c_{0}}{2}+\frac{a_0 c_{0}}{5} +\frac{a_1 c_{0}}{4} -\frac{9 a_0 c_{0}^2}{20}=0,
\end{eqnarray}
from where
\begin{equation} \label{a1}
a_1=2-\frac{4}{c_{0}}-\frac{4a_0}{5}+\frac{9a_0 c_{0}}{5}. 
\end{equation}
Using (\ref{a1}) in (\ref{I}) we obtain
\begin{eqnarray} \label{qa}
\frac{I}{r_0}&=&\frac{4}{c_{0}}-2+\frac{3a_0}{10}+\frac{2a_0 c_{0}}{5}+\ln \bigg( 1+\frac{1}{c_{0}}  \bigg) \bigg( -\frac{3}{c_0} \nonumber\\
&-&2  +c_0-\frac{2a_0 c_0}{5} -\frac{2a_0c_0^2}{5}  \bigg).
\end{eqnarray}

Next, replacing \eqref{a1} in (\ref{e11}) leads to
\begin{equation} \label{bdef}
    b(r)=\frac{2r_0}{c_{0}} +\bigg( 1-\frac{2}{c_{0}}-\frac{2a_0}{5} \bigg) \frac{r_0^2}{r}  +\frac{2a_0}{5}\frac{r_0^3}{r^2} ,
\end{equation}
with
\begin{eqnarray} 
&&c_{0}>0\label{e24}\\
&&a_0>\frac{5}{c_{0}}-5.\label{e25}
\end{eqnarray}

Then, as we require the minimum amount of exotic matter, we seek for the parameters that lead to a vanishing quantifier. To be more precise, we impose $I=0$ in (\ref{qa}) to obtain
\begin{equation} \label{a0}
    a_0=\frac{10 \big[4-2 c_{0}+
\ln\left(1+\frac{1}{c_{0}}\right)(c_{0}^{2}-2c_{0}-3)
 \big]}{ c_{0} \big[-3-4c_{0} +4c_0(c_0+1) \ln\big(1+\frac{1}{c_{0}}\big)\big]}.
\end{equation}

Finally, as  $b(r)>0$ for $r\in[r_{0},\infty)$, the only allowed roots are those appearing either at $r_{root}<r_{0}$ or at $r_{root}\in\mathbb{C}$.
From (\ref{bdef}), the roots of $b(r)$ are located at
\begin{equation} \label{rootg}
   r_{root}=\frac{-\beta\pm \sqrt{\beta^2-4\alpha\gamma}}{2\gamma},
\end{equation}
with
\begin{eqnarray}
\alpha&=&\frac{2a_0 r_0^3}{5}\\
\beta&=& \left(1-\frac{2}{c_{0}}-\frac{2a_0}{5}\right)r_{0}^{2}\\
\gamma&=&\frac{2r_0}{c_{0}}.
\end{eqnarray}

\begin{figure}[ht!] 
    \centering
    \includegraphics [scale=0.8] {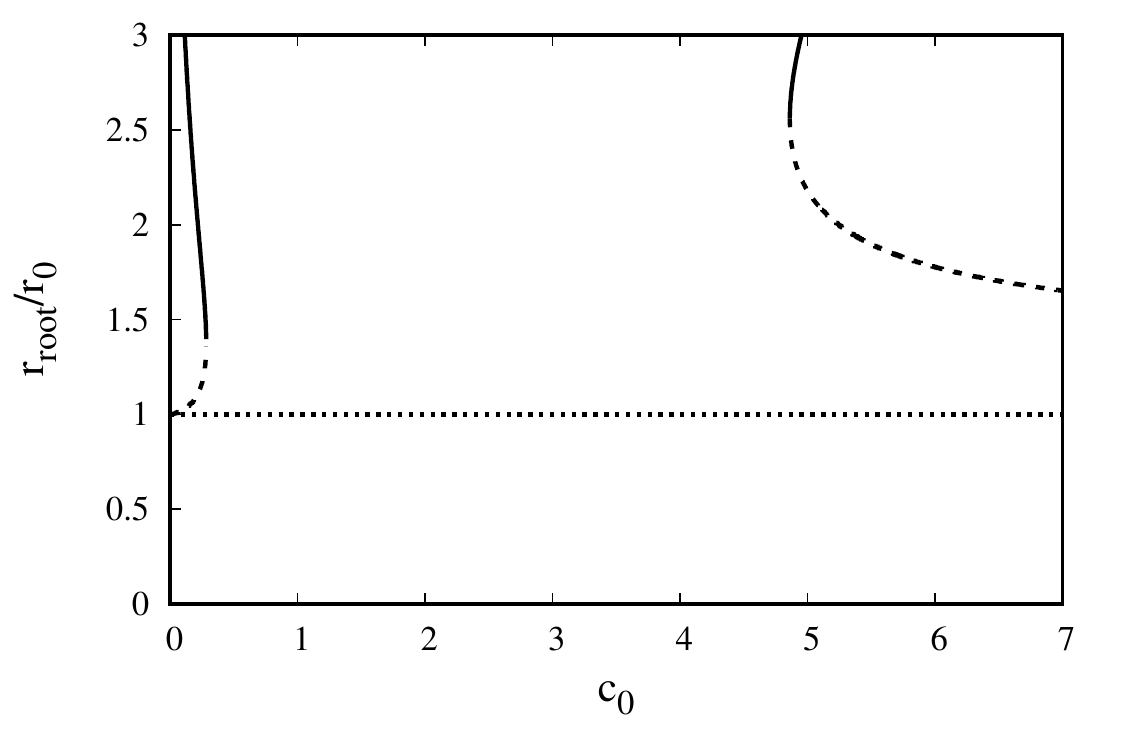}
    \caption{Real roots  of $b(r)$ as a function of $c_0$. The bigger roots correspond to the solid line and the smaller roots to the dashed line. The constant value of $1$ (dotted line) corresponds to the case $r_{root}=r_{0}$.}
    \label{fig:roots}
\end{figure}

It can be shown that whenever the roots are real,  $r_{root}>r_{0}$ (see Fig. \ref{fig:roots}).
In this respect, we must demand $\beta^2-4\alpha\gamma<0$ in order to ensure complex roots, which leads to
\begin{equation} \label{rootb}
   \bigg(1-\frac{2}{c_{0}}-\frac{2a_0}{5} \bigg)^2 -\frac{16a_0}{5c_{0}}<0,
\end{equation}
from where, by replacing $a_0$ given by \eqref{a0} we obtain that $c_{0}$
is restricted to values in the interval
\begin{equation} \label{brange}
   0.283181<c_{0}<4.86215.
\end{equation}
In summary, we have found that the only degree of freedom of a traversable wormhole with a minimum exotic matter, and complexity given by Eq. (\ref{e4}), is bounded as shown in (\ref{brange}). 

\begin{figure}[h!] 
    \centering
    \includegraphics [scale=0.8] {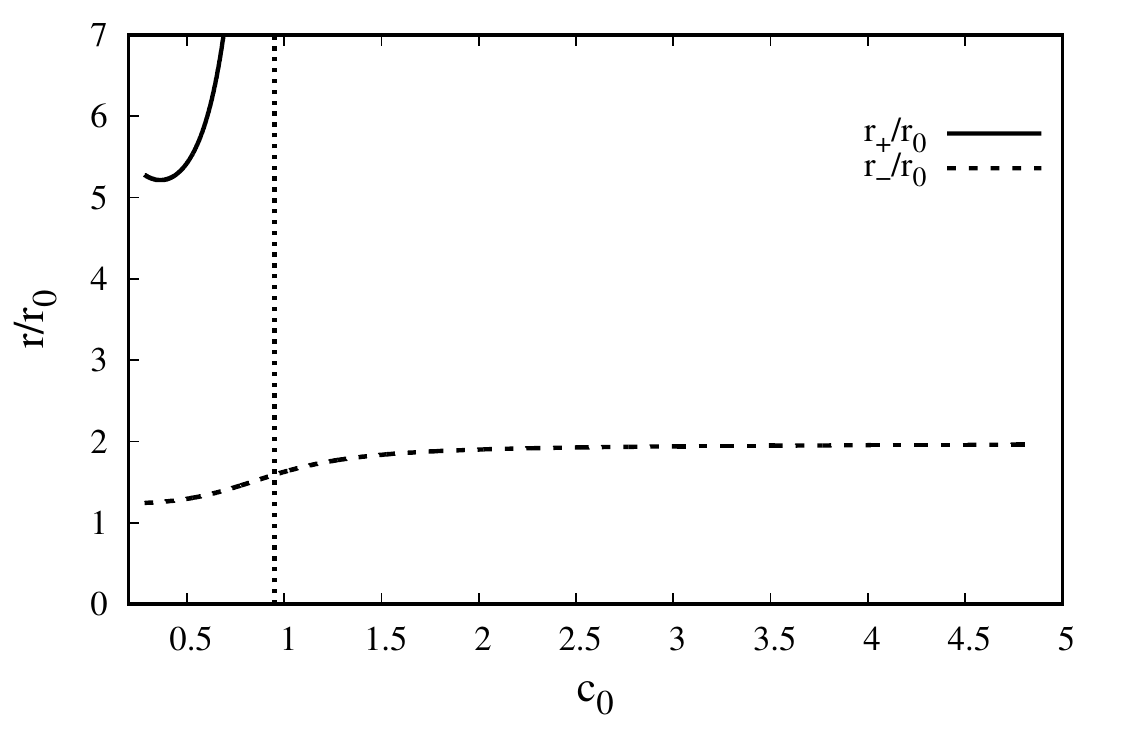}
    \caption{Normalized real positive roots  of $\rho +p_r$ as a function of $c_0$. The bigger roots $r_+/r_0$ correspond to the solid line and the smaller roots $r_-/r_0$ to the dashed line. The dotted line corresponds to the asymptotic limit of $r_+$ with value $c_0=0.950679$.}
    \label{fig:NECroots}
\end{figure}

In order to restrict the values of $c_{0}$ further, we can analyze the regions where the NEC is violated which, as we stated before, must be located near the throat of the wormhole as a consequence of the flaring--out condition. In this regard,  as $\rho+p_{r}<0$ near the throat, we require that such a quantity change of sign at certain $r\approx r_{0}$ and remains positive everywhere in
$r\in (r_{0},\infty)$. In Fig. \ref{fig:NECroots} we show the roots of $\rho+p_{r}$ and observe that as $c_{0}$ approach to $0.950679$ from below, the NEC is not only violated near the horizon (for $r\in(r_{0},r_{-}$) but in regions far from $r_{0}$ (for $r\in(r_{+},\infty)$) .
Consequently, in order to avoid the violation of NEC far from $r_{0}$ we require
 \begin{eqnarray}\label{internew}
 0.950679<c_{0}<4.86215.
 \end{eqnarray}
 
\begin{figure}[h!] 
    \centering
    \includegraphics [scale=0.8] {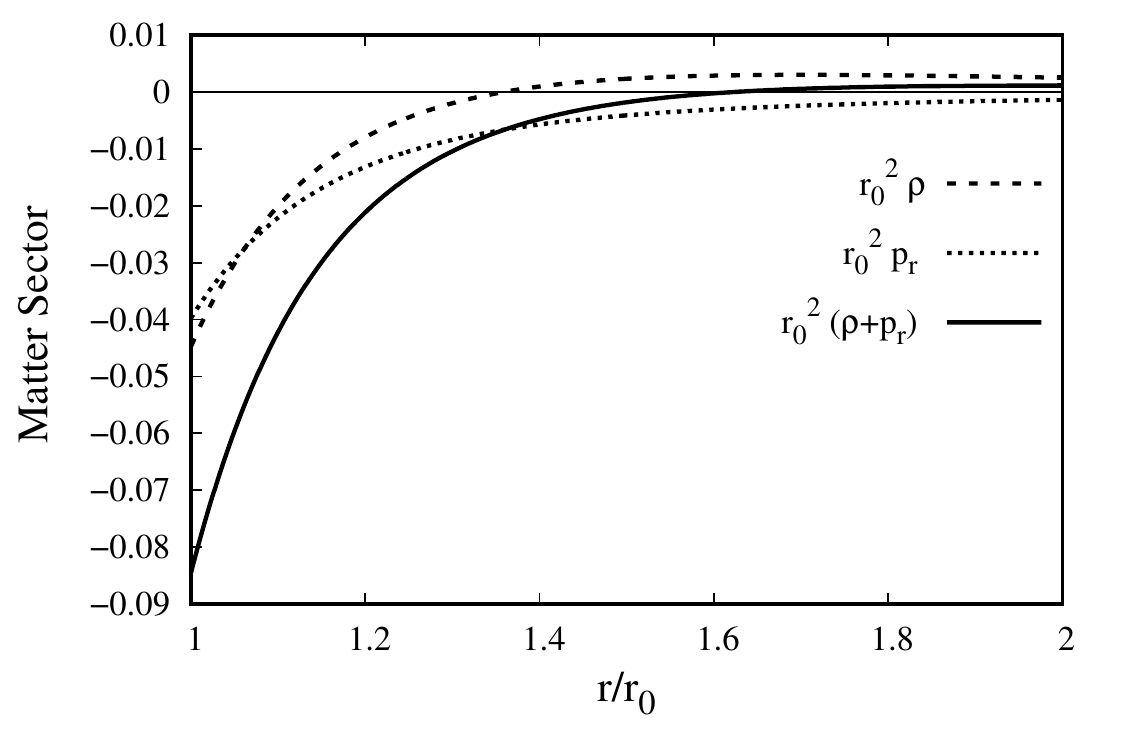}
    \caption{Normalized density $r_0^2 \rho$ (dashed line), normalized radial pressure $r_0^2 p_r$ (dotted line), and normalized density plus radial pressure $r_0^2 (\rho + pr)$ (solid line) as a function of $r/r_0$. Fixed parameter $c_0=1$.}
    \label{fig:matter}
\end{figure}

To complement the discussion on the matter sector supporting the wormhole geometry, in Fig \ref{fig:matter} we show the profiles of the energy density, $\rho$, the radial pressure, $p_{r}$, and $\rho+p_{r}$ for the parameters in the legend \footnote{We do not show $\rho+p_{t}$ because this quantity is positive everywhere for the values of $c_{0}$ in (\ref{internew}).
}.
Note that both, the energy density and the radial pressure, have a like--Casimir behaviour near the throat in the sense that are negative as reported in \cite{remo}. Of course, such a similarity must be taken as merely formal in the sense that, in this case, the equation of state of the matter sector can be written 
\begin{eqnarray}
p_r = \omega \rho ,\\
p_t = \omega_{t}\rho ,
\label{EoS}
\end{eqnarray}
with 
\begin{eqnarray}
\omega&=&\frac{\alpha_1 r^2 + \beta_1 r + \gamma_1}{(c_0 r +r_0) (\mu _1 r +\nu_1)}\\
\omega_{t}&=&\frac{\alpha_2 r^3 + \beta_2 r^2 + \gamma_2 r +\delta_2}{(c_0 r +r_0)^2 (\mu _2 r +\nu_2)}, 
\end{eqnarray}
where the coefficients $\alpha_1$, $\beta_1$, $\gamma_1$, $\mu_1$, $\nu_1$, $\alpha_2$, $\beta_2$, $\gamma_2$, $\delta_2$, $\mu_2$, and $\nu_2$ depend of the coefficients $c_0$ and $r_0$. Besides, in contrast to the traversable wormhole in \cite{remo}, the like--Casimir behaviour is bounded in a region near the throat and not in the hole spacetime which requires that $\rho<0$ everywhere. In this regard, we could say that there is a kind of ``phase transition'' at certain $r$ where the matter goes from a content with quantum properties enclosed near the throat (where it is required) and classical matter in the rest of the space--time. Furthermore, we could conjecture that such a behaviour is only possible whenever the quantifier vanishes exactly (as demanded here).

From now on we shall explore the features of the traversable wormhole geometry constructed here based on the values of $c_{0}$ in (\ref{internew}).

\begin{figure}[h!]
    \centering
    \includegraphics [scale=0.8] {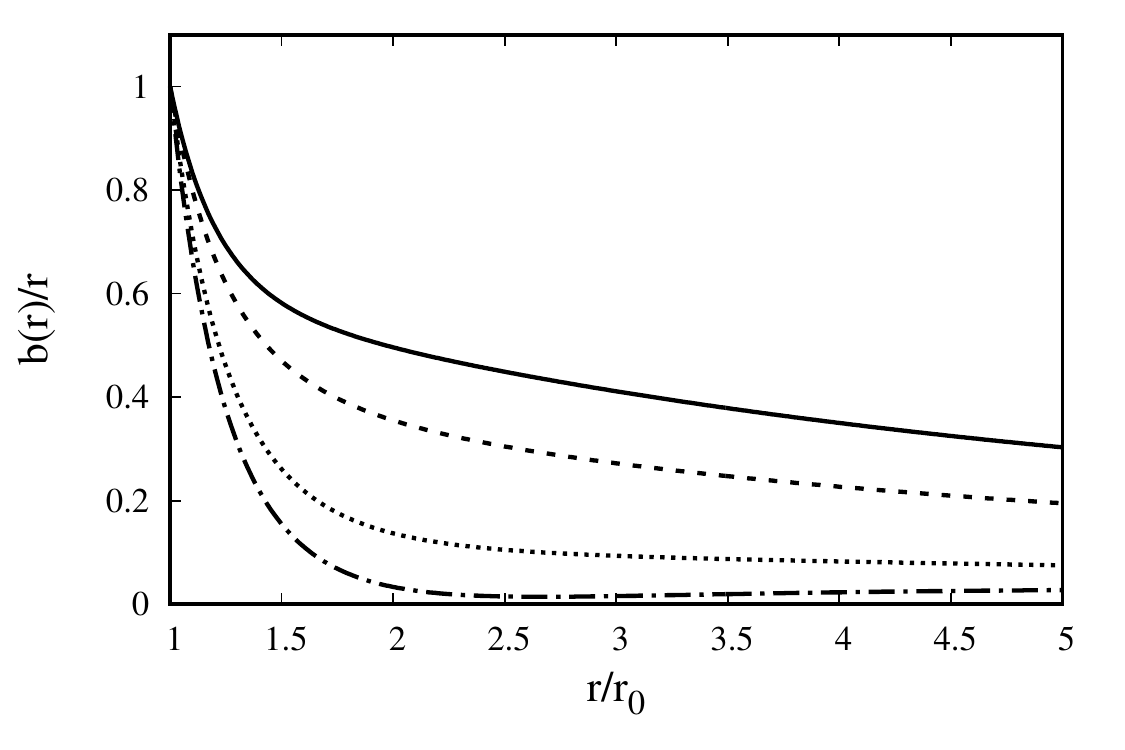}
    \caption{$b/r$ as a function of $r/r_{0}$  for $c_{0}=0.96$ (solid line), $c_{0}=1.5$ (dashed line), $c_{0}=3$ (dotted line), $c_{0}=4.5$ (dash-dotted line).}
    \label{fig:shaperQ0}
\end{figure}

\begin{figure}[h!]
    \centering
    \includegraphics [scale=0.8] {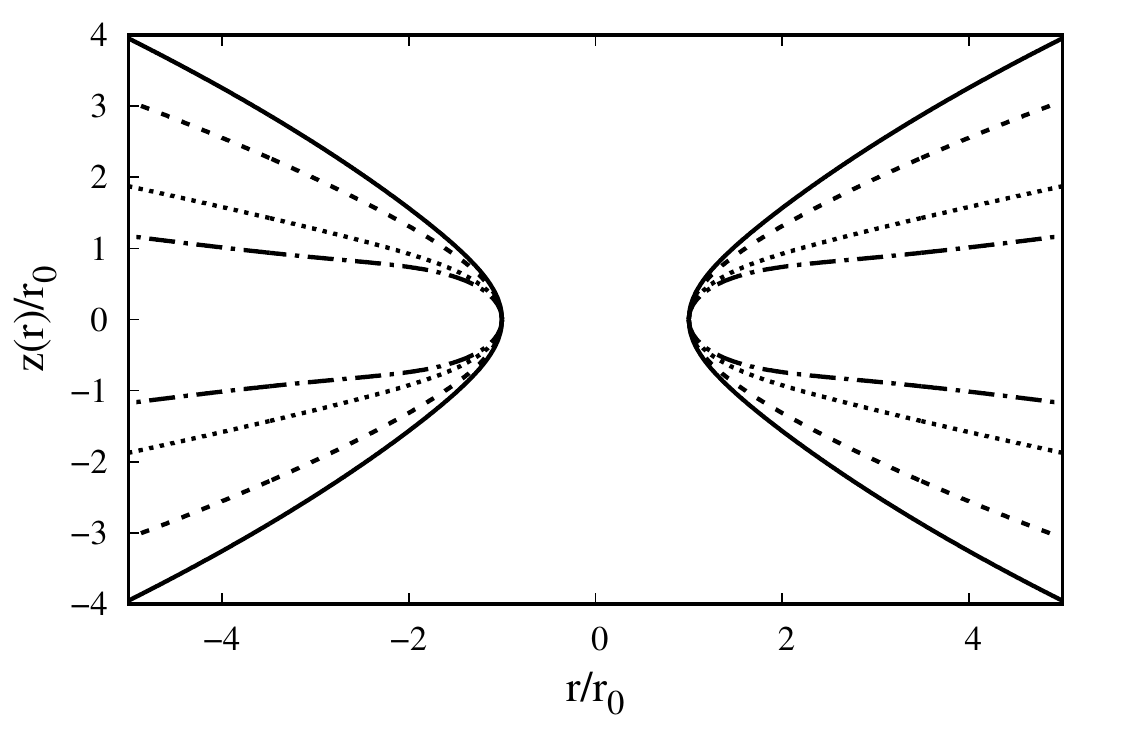}
    \caption{Embedding diagram  for $c_{0}=0.96$ (Solid line), $c_{0}=1.5$ (Dashed line), $c_{0}=3$ (Dotted line), $c_{0}=4.5$ (Dash-dotted line).}
    \label{fig:embQ0}
\end{figure}

In Fig. \ref{fig:shaperQ0} we show $b/r$ as a function of the radial coordinate for the choice of parameters shown in the legend of the figure. Note that the speed of convergence increases as $c_{0}$ grows. In  Fig. \ref{fig:embQ0} it is shown the embedding diagram (for the parameters displayed in the legend) for the traversable wormhole. Notice that $dz/dr\to0$ as $r\to\infty$ goes faster as $c_{0}$ increases. 

\begin{figure}[ht!] 
    \centering
    \includegraphics [scale=0.8] {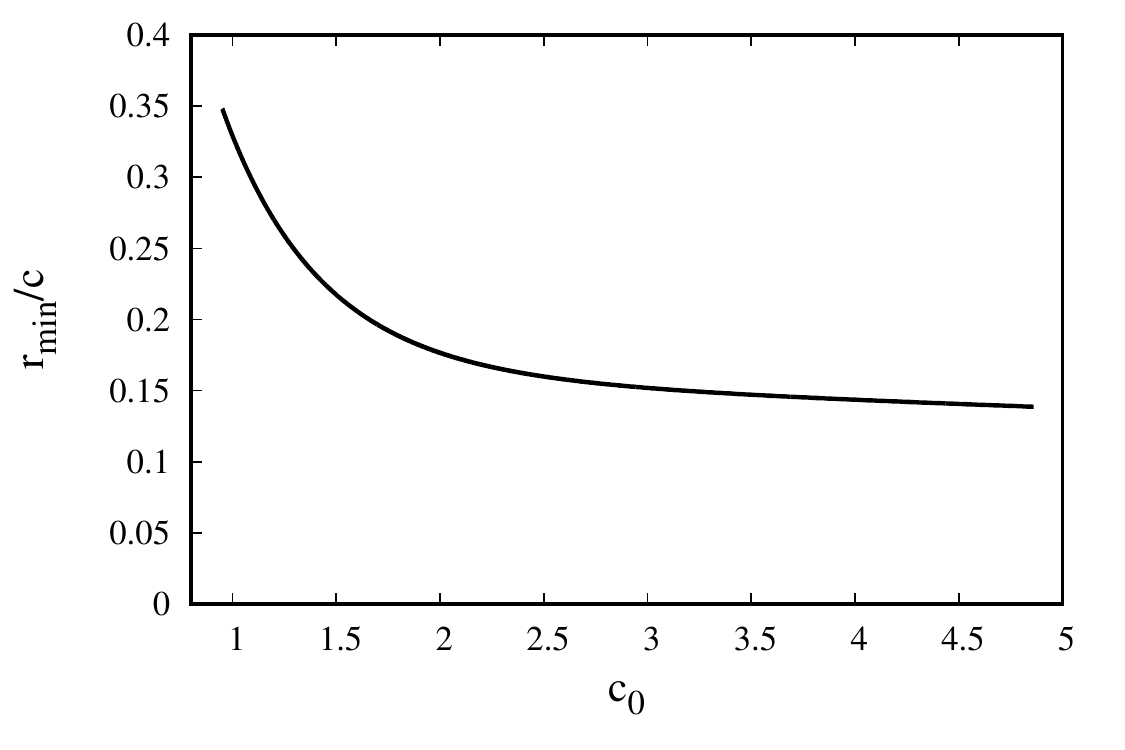}
    \caption{Minimum radius of the wormhole throat $r_{min}$ normalized by $c$ as a function of $c_0$.}
    \label{fig:rmin}
\end{figure}

\begin{figure}[ht!]
    \centering
    \includegraphics [scale=0.8] {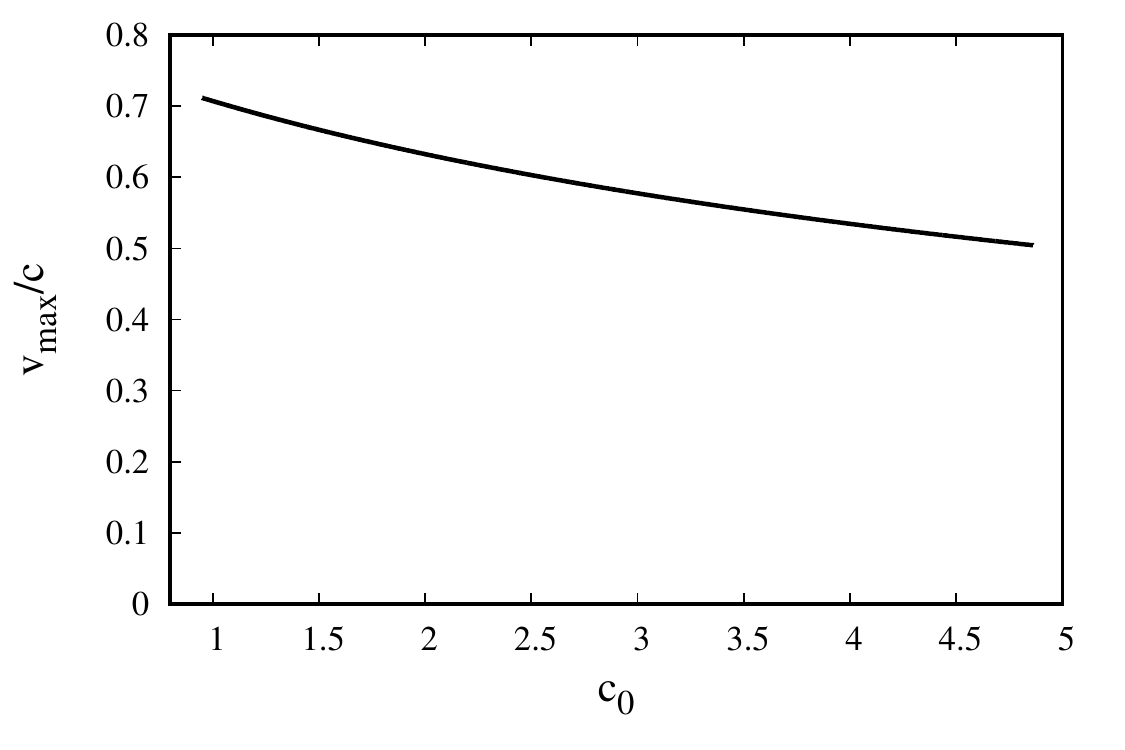}
    \caption{Maximum velocity of the traveler $v_{max}$ normalized by the speed of light $c$  as a function of $c_0$.}
    \label{fig:vmax}
\end{figure}

It is worth mentioning that, as our wormhole fulfills the basic requirement to be traversable, it should be interesting to explore if this geometry could allow the human interstellar travel. To this end we must ensure that the tidal accelerations experienced by the traveller must be of the order (or less) than the Earth's gravitational acceleration. In this regard, we shall use 
Eqs. (\ref{tidalr}) and (\ref{tidalt})
to estimate the size of the throat and the velocity of the traveller at the throat (assumed as constant for simplicity). At the throat, Eqs. (\ref{tidalr}) and (\ref{tidalt}) can be written as
\begin{eqnarray}
&& |\phi ' (r_0)|\le \frac{2g_{\oplus}r_0^2}{(1-b'(r_0))|\eta ^1|c^2}\\
&&\gamma^2 v^2\le \frac{2g_{\oplus }r_0}{(1-b'(r_0))|\eta ^2|c^2},
\end{eqnarray}
 which leads to 
 \begin{eqnarray}
&&  \sqrt{\frac{|\eta^1| \xi_1}{g_\oplus (1+c_0)}} c\le r_0 \label{rmin}\\ 
&&v\le \sqrt{\frac{\xi_2}{1+\xi_2}} c,  \label{vmax}
\end{eqnarray}
where $\xi_1=\big|1- \frac{1}{c_0}-\frac{a_0}{5} \big|$, $\xi_2=\frac{g_\oplus r_0^2}{|\eta^2| \xi_1 c^2}$, $a_0$ is given by Eq. \eqref{a0}
and $\eta^1\approx 2[m]$ and $\eta^2\approx 1[m]$ as the usual dimensions for a person. Notice that when inequalities (\ref{rmin}) and (\ref{vmax}) are saturated they imply a minimum value for $r_0$ and a maximum value for $v$. The minimum value for $r_0$ is only dependent on the parameter $c_0$ and can be written as
\begin{eqnarray}
r_{min}=\sqrt{\frac{|\eta^1| \xi_1}{g_\oplus (1+c_0)}} c,
\end{eqnarray}
so it depends on the constraint on $c_{0}$ given by (\ref{brange}). In Fig. \ref{fig:rmin} it is shown the behaviour
of $r_{min}/c$ as a function of $c_{0}$.
Note that, 
\begin{eqnarray}
0.41\times 10^8 [m] \le r_{min} \le 1.04\times 10^8 [m], 
\end{eqnarray}
which is of the same order of the Earth-Moon distance.
Now, in order to estimate the maximum velocity we shall use $r_0=r_{min}$ form where \eqref{vmax} reads
\begin{eqnarray}
v_{max}=\sqrt{\frac{|\eta^1|}{|\eta^2|(1+c_0)+|\eta^1|}}c \label{vmaxc0}.
\end{eqnarray}
In Fig. \ref{fig:vmax} we show  $v_{max}$ as a function of $c_{0}$. In this case we obtain \begin{eqnarray}
1.51\times 10^8 [m/s]\le v_{\max}\le 2.13\times 10^8 [m/s],
\end{eqnarray}
which is less than the speed of light.

\begin{figure}[ht!] 
    \centering
    \includegraphics [scale=0.8] {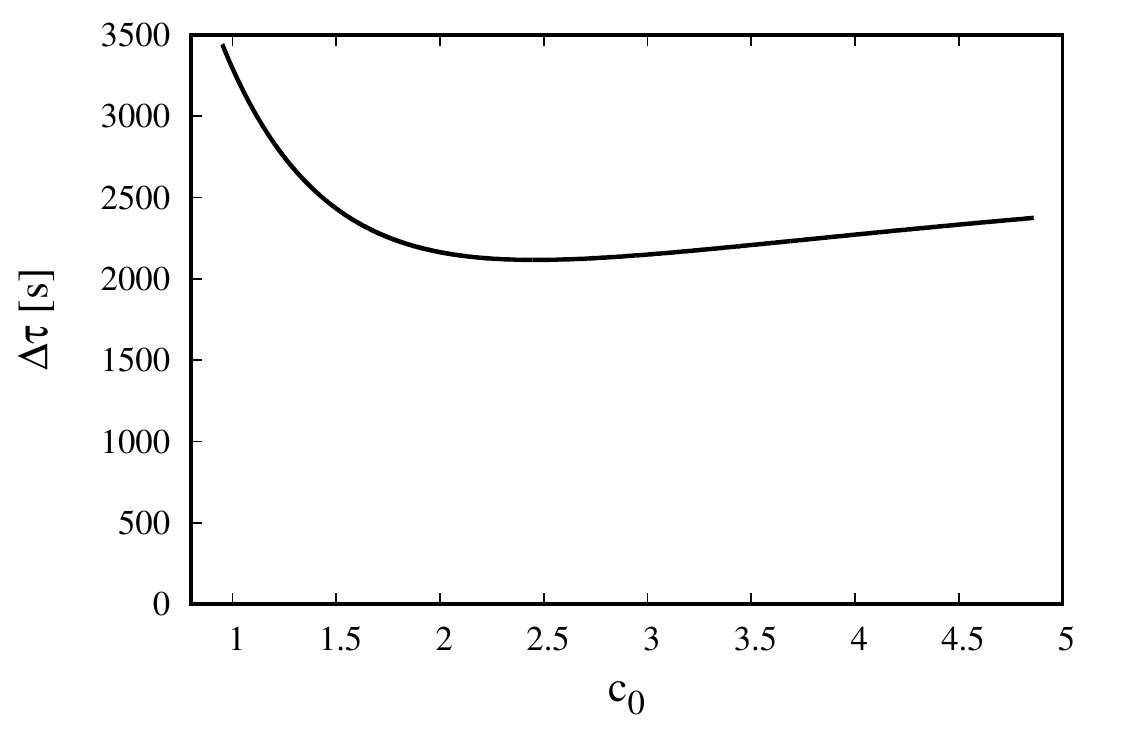}
    \caption{Coordinate time $\Delta t$ as a function of $c_0$.}
    \label{fig:dt}
\end{figure}

\begin{figure}[ht!]
    \centering
    \includegraphics [scale=0.8] {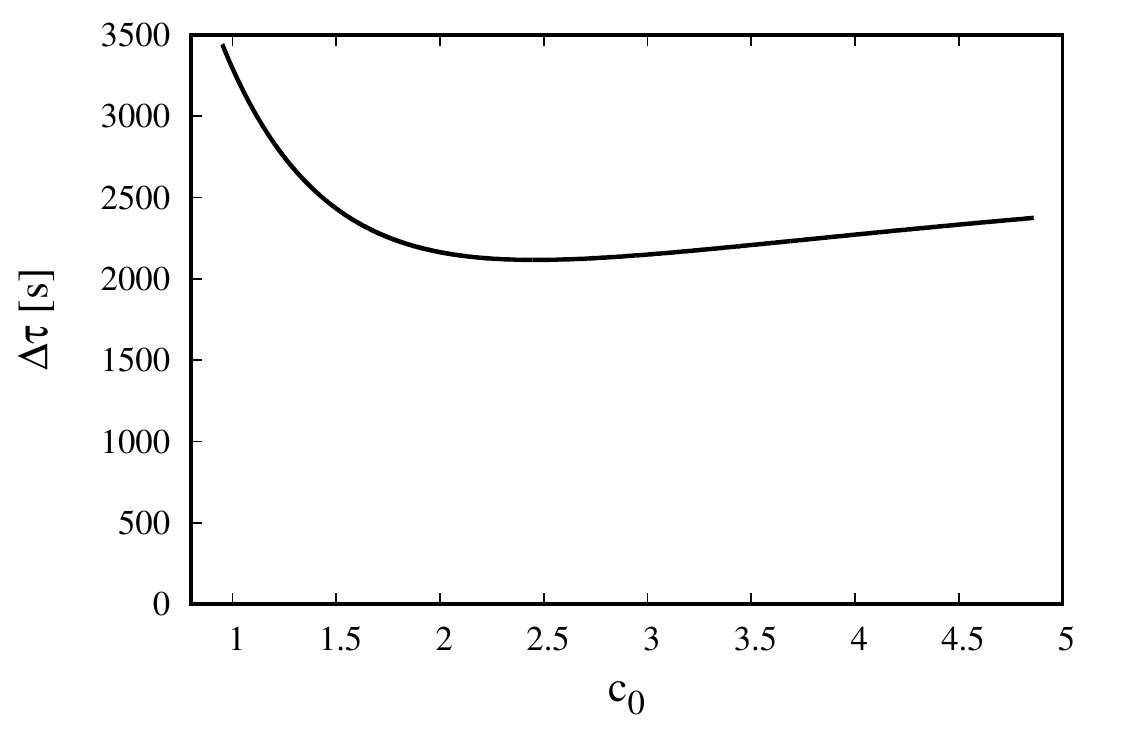}
    \caption{Proper time $\Delta \tau$ as a function of $c_0$.}
    \label{fig:dT}
\end{figure}

Finally, as stated previously, a reasonably interstellar travel  should take no more than a year. Here we estimate the value of both the coordinate and the proper time from
Eqs. (\ref{coort}) and (\ref{propt}) by setting the distance of the spatial stations (located at each asymptotically flat region) $r_{st}=10^{4}r_{0}$. The results are depicted in Figs.
(\ref{fig:dt}) and (\ref{fig:dT})
where is shown that both, the maximum coordinate and the proper times are of the order of $10^{3}s$ which corresponds to a total time around $1h$.

\section{Final remarks}\label{concl}

In this work we implemented the recently introduced concept of complexity to construct wormhole geometries belonging to the same ``equivalence class'' of the Casimir wormhole reported in \cite{remo}. The solution could be characterized with only one  parameter appearing as a consequence of the generalization of the redshift function of the Casimir wormhole \cite{remo}. The values of such a parameter were constrained by demanding the basic properties that a wormhole geometry has to fulfill in order to be traversable; namely i) the flaring-out condition, ii) tidal acceleration of the order of the Earth gravitational acceleration iii) finite time to travel from a spatial station to the throat and iv) a minimal amount of exotic matter. As a consequence we found that the solution connects two asymptotically flat regions through a tunnel with the size of the Earth-Moon distance and the time required to traverse the wormhole from a spatial station located in the asymptotically flat region it is on the order of a few hours. Probably, the most intriguing feature (in contrast to the solution in \cite{remo}) is that the quantifier does not depends on the size of the throat but is arbitrarily small which means that, we can construct the traversable wormhole with a minimal quantity of exotic matter. Another interesting feature is that, in contrast to \cite{remo}, the like--Casimir behaviour of the matter sector obtained here (in the sense that both the energy density and the radial pressure are negative) is bounded by a radius near the throat such that, for bigger radial distances, the matter sector is completely classical (density and pressure positives). This represent an advantage because, in particular, the energy density is negative only in a small region and not the whole space--time as in \cite{remo}.

Before concluding this work, we would like to address a couple of points that we think deserve special attention. First, we could consider 
\begin{eqnarray}\label{eso1}
\frac{a_{5}r_{0}b'(r_{0}-3r_{0}a_{6})}{2r^{3}}
\end{eqnarray}
instead
\begin{eqnarray}\label{eso2}
\frac{r_{0}b'(r_{0}-3r_{0})}{2r^{3}}
\end{eqnarray}
as an extra term in Eq. (\ref{e4}). Clearly, (\ref{eso1}) reduces to (\ref{eso2}) when $a_{5}=a_{6}=1$ but for arbitrary values it could lead to a shape function which would generalize the one obtained here and would enrich the discussion (we acknowledge the anonymous referee for drawing our attention on this point). Second, it should be interesting the study of the stability of the wormhole geometry obtained here. However, these and other issues are out of the scope of this work and shall be considered in future developments.

\section{Acknowledgements}
E.C  acknowledges  Decanato  de  Investigaci\'on  y  Creatividad, USFQ, Ecuador, for continuous support. The authors acknowledge the anonymous referee for his/her valuable comments which enriched the discussion of the work.

\end{document}